\documentclass{ws-ijmpa}
\usepackage[super,compress]{cite}
\usepackage{graphicx}

\usepackage{xurl}
\usepackage{url}
\usepackage{hyperref}
\begin{document}
\markboth{R. Manikandhan}{}

%
\catchline{}{}{}{}{}
%

\title{Recent highlights from the STAR Experiment}

\author{Rutik Manikandhan, for the STAR Collaboration}

\address{Department of Physics, University of Houston, Houston, TX 77204, USA\\
manikandhan.rutik@gmail.com}

\maketitle

\begin{history}
\received{$8^{th}$ July 2026}
\end{history}

\begin{abstract}
Understanding the QCD phase structure and the possible existence of a critical point remains one of the central goals of the heavy-ion program at RHIC. In this talk, we will present recent STAR results across multiple observables that probe different aspects of the hot and dense matter created in Au+Au collisions. 
These include two-particle transverse momentum correlations of $ \langle p_{T} \rangle $, net-proton cumulants up to fourth order, identical pion femtoscopy, and baryon-strangeness correlations. We will also discuss femtoscopic measurements of baryon-baryon pairs which offer insight into hyperon-nucleon and hyperon-hyperon interactions and the possible formation of strange dibaryon states. Together, these results provide complementary probes of the system’s evolution across a wide energy range ($\sqrt{s_{NN}}$ = 3–200 GeV), offering new constraints on the QCD equation of state and the location of the critical point.

\keywords{Heavy-ion Collisions; Beam Energy Scan; Critical point.}
\end{abstract}



\section{Introduction}	

Understanding the phase structure of Quantum Chromodynamics (QCD) at finite temperature and baryon density remains a central goal of the heavy-ion collision program at the Relativistic Heavy Ion Collider (RHIC). In particular, the possible existence of a critical point in the QCD phase diagram is expected to manifest through anomalous event-by-event fluctuations and long-range correlations in conserved quantities as well as in the collective behavior of the produced medium~\cite{PhysRevD.60.114028,Stephanov_1998}.
A rich variety of observables have been proposed to probe different aspects of the hot and dense QCD matter. Higher-order cumulants of net-proton multiplicity distributions are sensitive to the correlation length of the system and are predicted to show non-monotonic behavior near the critical point~\cite{PhysRevLett.107.052301,STAR:2021netproton}. Event-by-event fluctuations and two-particle correlations of transverse momentum ($  p_T  $) provide insight into the nuclear structure~\cite{STAR:2024wgy,STAR:2025elk} and can serve as a probe of critical fluctuations by studying the extent of thermalization in the system~\cite{PhysRevC.85.014905,PhysRevLett.92.162301}. Femtoscopic measurements, both for identical pions and baryon-baryon pairs, offer unique access to the space-time evolution of the system and the final-state interactions between hadrons. In addition, correlations between baryon number and strangeness provide valuable information on the nature of the degrees of freedom~\cite{PhysRevLett.95.182301} in the medium. 
The Beam Energy Scan (BES) program at RHIC, and in particular the recent BES-II data, offers an ideal dataset to search for these signatures by varying the collision energy from $  \sqrt{s_{NN}} = 200~$GeV down to $3$~GeV, thereby scanning a wide range of the QCD phase diagram~\cite{Chen:2024aom}.
In these proceedings, we present recent STAR results on multiple complementary observables measured in Au+Au collisions. These include two-particle transverse momentum correlations and event-by-event fluctuations of $\langle p_{\rm T} \rangle$, net-proton cumulants up to fourth order, identical pion femtoscopy, baryon-strangeness correlations, and femtoscopic studies of baryon-baryon pairs. Together, these measurements provide new constraints on the QCD equation of state and offer fresh insight into the possible location of the critical point in the QCD phase diagram.

\section{Results}

\subsection{Net-proton Cumulants}
The results shown here for the proton cumulants are within a common kinematic acceptance within a rapidity window $(|y| < 0.5)$ across all collider energies. The Time-Projection Chamber (TPC) and Time-Of-Flight detectors have been used for identifying the protons and the anti-protons. The TPC identifies the low $p_{T}$ ($0.4 < p_{T} < 0.8 $~GeV/c ) protons and anti-protons with high purity and the TOF identifies particles at higher $p_{\rm T}$ ($0.8 < p_{\rm T} < 2.0$~GeV/c ). For the fixed-target energies at nucleon-nucleon center of mass energies of $\sqrt{s_{NN}}$ = 3.0, 3.2, 3.5 $\&$ 3.9~GeV due to limited acceptance\cite{STAR:2023protoncumulants}  we show results for $ -0.5 <y-y_{cm} <0$.

\begin{figure}[ht]
\centerline{\includegraphics[width=12.5cm]{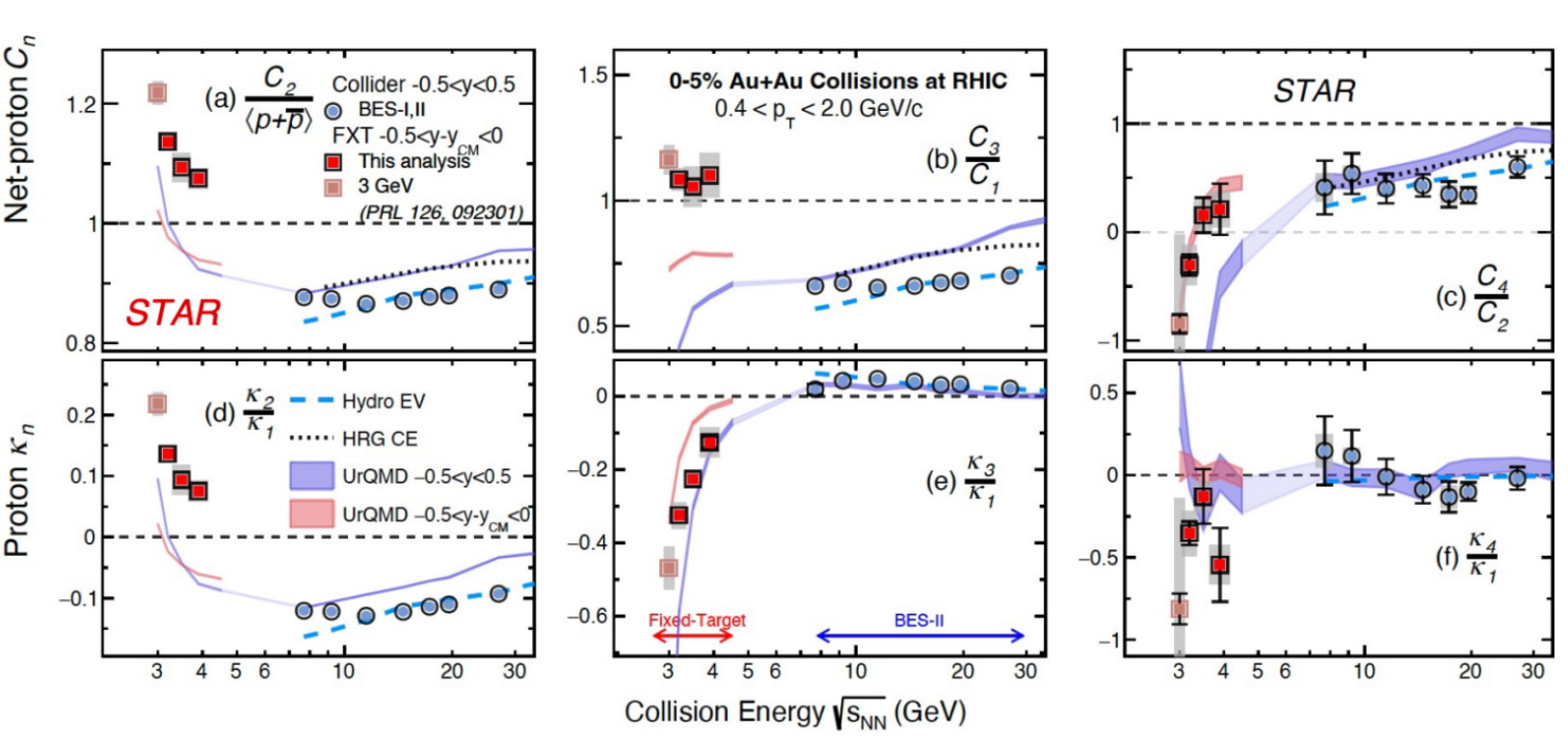}}
\caption{Net-proton cumulant ratios: (a) $C_2/\langle p+\bar{p}\rangle$, (b) $C_3/C_1$, and (c) $C_4/C_2$, and proton factorial cumulant ratios: (d) $\kappa_2/\kappa_1$, (e) $\kappa_3/\kappa_1$, and (f) $\kappa_4/\kappa_1$ in Au+Au collisions. The bars and bands on the data points represent statistical and systematic uncertainties, respectively. Theoretical calculations from a hydrodynamical model with excluded volume~\cite{PhysRevC.105.014904} (Hydro, blue dashed line), a thermal model with canonical treatment for baryon charge~\cite{Braun_Munzinger_2021} (HRG CE, black dashed line), a transport model~\cite{Bass_1998,Bleicher_1999} (UrQMD, blue band and red band). Figure taken from Ref.~\cite{ZSQM2025} \label{f1}}
\end{figure}

The collider-energy measurements of $C_{2}/C_{1}$ and $\kappa_{2}/\kappa_{1}$ exhibit a monotonic increase with collision energy, consistent with hydrodynamic model expectations. In contrast, the fixed-target results show a monotonic decrease in both observables as the collision energy decreases, in qualitative agreement with UrQMD predictions. In addition, at fixed-target energies we observe a significant enhancement of the cumulants relative to baseline expectations from transport (UrQMD~\cite{Bleicher_1999}), hydrodynamics with excluded volume~\cite{Braun_Munzinger_2021}, and thermal-model calculations with a canonical treatment of baryon number~\cite{PhysRevC.105.014904}.

The $C_{3}/C_{1}$ observable does not exhibit a strong dependence on collision energy and shows behavior qualitatively similar to the UrQMD expectations. In contrast, $\kappa_{3}/\kappa_{1}$ displays a monotonic increase with collision energy, consistent with the predicted trend. At the fixed-target energies, significant deviations from the non-critical baseline are observed in both $C_{3}/C_{1}$ and $\kappa_{3}/\kappa_{1}$. These deviations may indicate enhanced dynamical contributions in the low-energy region and motivate further investigation of higher-order fluctuation observables.

The $C_{4}/C_{2}$ measurements at the fixed-target energies remain consistent with the baseline expectations within uncertainties. The largest deviation from the baseline is observed at $\sqrt{s_{NN}} = 19.6$~GeV, with a significance of approximately $2$--$5\sigma$. No strong collision-energy dependence is observed for $\kappa_{4}/\kappa_{1}$ within the current experimental uncertainties.

\subsection{Transverse Momentum Correlations}
Two-particle $p_{\rm T}$ correlations are studied using the scaled correlator $\sqrt{\langle \Delta p_{t,i} \Delta p_{t,j} \rangle}/\langle \langle p_{\rm T} \rangle \rangle$. It represents the magnitude of the dynamic fluctuations of the average transverse momentum in units of $\langle \langle p_{\rm T} \rangle \rangle$. This scaling cancels out detector efficiency effects \cite{STAR:2019dow,ALICE:2014gvd} and flow effects \cite{PhysRevLett.92.162301} making it an ideal probe for critical point searches.

\begin{figure}[ht!]
\centerline{\includegraphics[width=9.5cm]{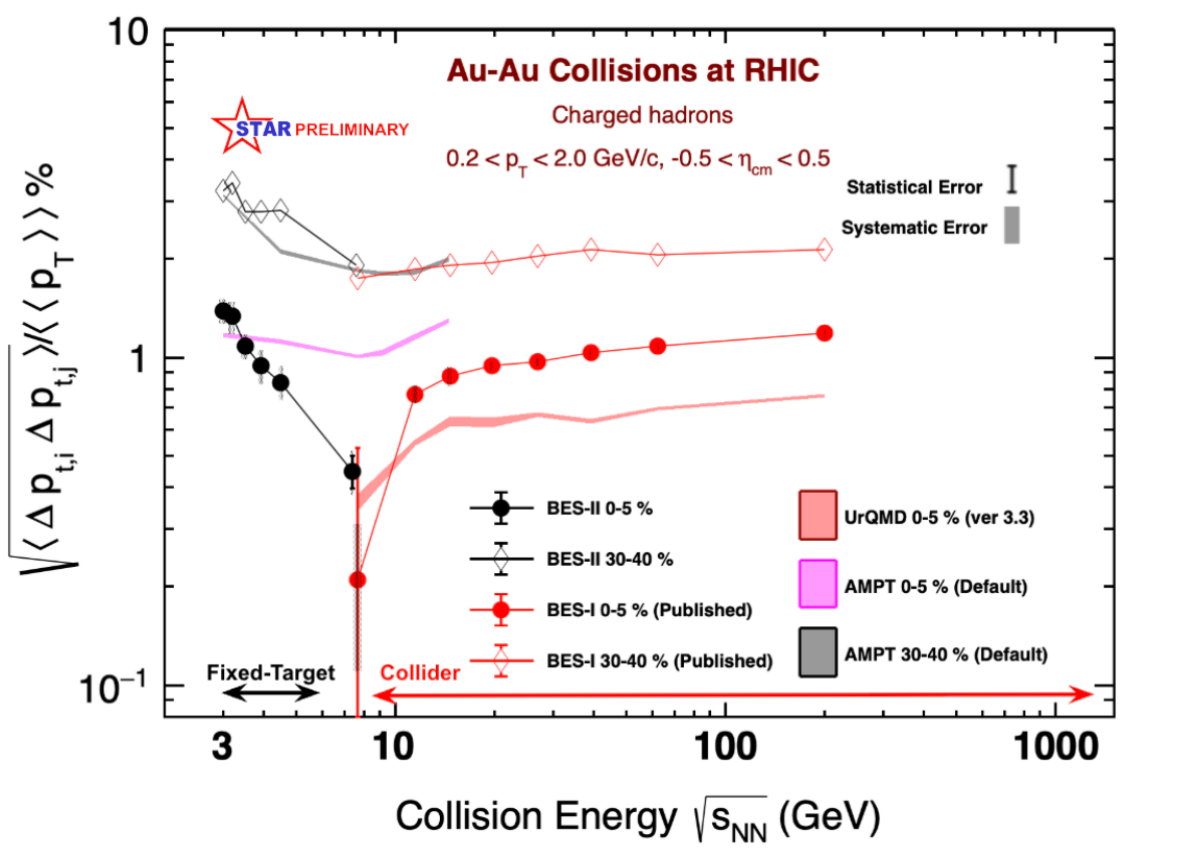}}
\caption{ The relative dynamical correlation in Au+Au 0–5\% central collisions from this analysis, together with BES I data. Statistical and systematic uncertainties are shown as bars and gray shading, respectively. For all measurements, charged particles are selected within a $p_{\rm T}$ acceptance of [0.2, 2.0] GeV/c and a pseudorapidity ($\eta$) acceptance of $|\eta_{\rm cm}| < 0.5$, where $\eta_{\rm cm} = \eta_{\rm lab} - \eta_{\rm mid}$. AMPT~\cite{PhysRevC.111.024911} (0-5\%) and (30–40\%) shown as pink (grey) bands, UrQMD ver. 3.3~\cite{Bleicher_1999} calculations are shown as red bands. Figure taken from Ref.~\cite{RMQM2025}.\label{f2}}
\end{figure}

In Fig.~\ref{f2}, a pronounced non-monotonic dependence of the scaled correlator on collision energy is observed for the most central collisions. The structure becomes progressively weaker toward peripheral collisions, indicating a strong dependence on the size and lifetime of the created medium. In particular, the measurements exhibit a minimum in the intermediate collision-energy region, which may reflect changes in the underlying thermodynamic properties of the system.

The available transport model calculations (UrQMD~\cite{Bleicher_1999} and AMPT~\cite{PhysRevC.111.024911}) are unable to quantitatively reproduce the measured energy dependence, especially in the central collisions where the non-monotonic behavior is most pronounced. Since conventional hadronic transport models primarily incorporate non-critical contributions, the discrepancy may suggest the presence of additional physics mechanisms not fully captured in the current calculations.

The BES-II measurement at $\sqrt{s_{NN}} = 7.7$~GeV, obtained with significantly higher statistics, exhibits substantially reduced uncertainties while remaining consistent with the previously published BES-I result~\cite{STAR:2019dow}. This agreement demonstrates the robustness of the observable against differences in detector conditions and analysis procedures between the BES-I and BES-II data-taking periods, while also providing improved sensitivity to the observed non-monotonic trend.

\subsection{Baryon-Strangeness Correlations}

Correlations between conserved quantum numbers provide important insight into the microscopic degrees of freedom of strongly interacting matter. In particular, baryon--strangeness correlations have been proposed as a sensitive probe of the QCD medium~\cite{PhysRevLett.95.182301}. In an ideal quark--gluon plasma, strangeness is carried by strange quarks, which simultaneously carry baryon number, leading to a strong correlation between baryon number and strangeness. In contrast, in a hadron gas a significant fraction of the strangeness is carried by kaons, which possess zero baryon number, resulting in weaker correlations. Therefore, measurements of baryon--strangeness correlations can help distinguish between hadronic and partonic degrees of freedom and provide information on the QCD phase structure.

\begin{figure}[ht!]
\centerline{\includegraphics[width=11.5cm]{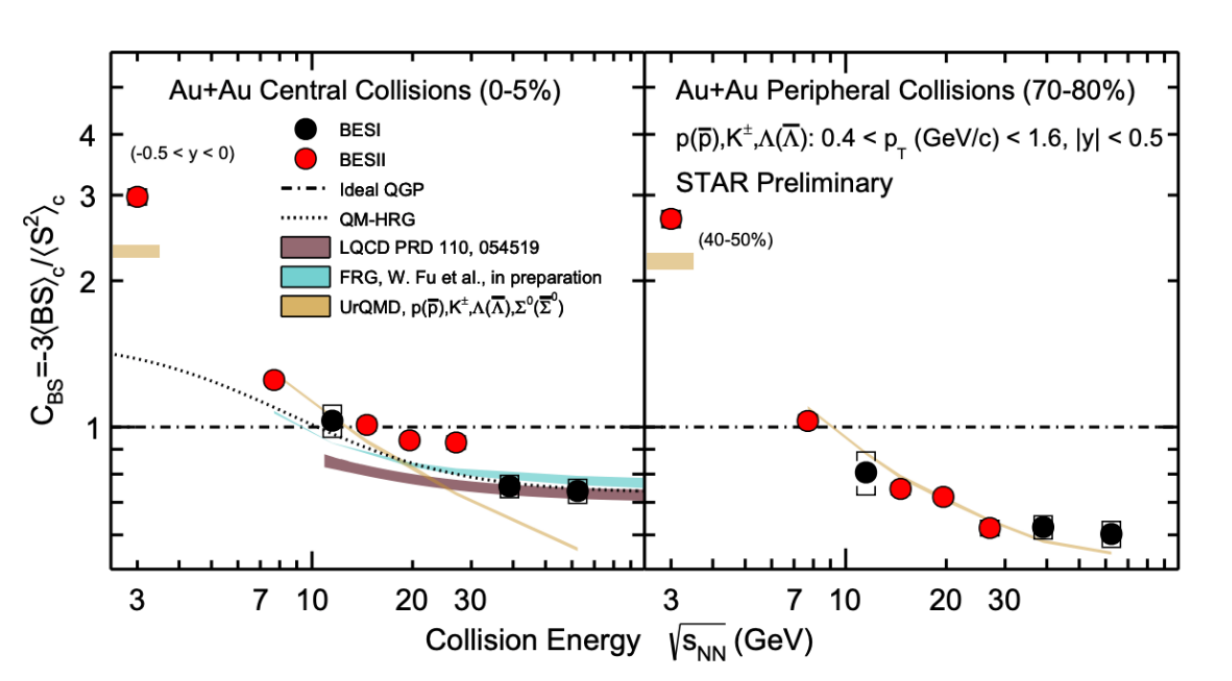}}
\caption{Collision-energy dependence of the baryon--strangeness correlation observable in Au+Au collisions at RHIC. The measurements are compared with theoretical calculations and expectations from hadronic (UrQMD) and partonic (FRG,LQCD) scenarios. Statistical uncertainties are shown as bars, while systematic uncertainties are represented by shaded boxes/bands where applicable. Figure taken from Ref.~\cite{HFQM2025}.\label{f3}}
\end{figure}

Figure~\ref{f3} presents the collision-energy dependence of the baryon--strangeness correlation observable measured in Au+Au collisions at RHIC. The measurements are compared with theoretical expectations from hadronic and partonic scenarios. A clear energy dependence is observed, indicating that the underlying correlation between baryon number and strangeness evolves significantly across the BES energy range.

At lower collision energies, where the baryon chemical potential is large, the measured correlations are consistent with expectations from a hadron-dominated medium from the UrQMD calculations. Toward higher collision energies, the observables approach values expected from deconfined quark degrees of freedom from the FRG and LQCD~\cite{PhysRevD.110.054519} calculations, suggesting an increasing role of partonic interactions. The measurements are qualitatively consistent with the picture proposed in Ref.~\cite{PhysRevLett.95.182301}, where baryon--strangeness correlations serve as a diagnostic of the active carriers of strangeness in the medium.

The comparison with model calculations indicates that no single model simultaneously reproduces the full energy dependence and magnitude of the measurements. These results provide important constraints on theoretical descriptions of the QCD medium and motivate further studies with higher precision measurements in the BES-II and future fixed-target programs.

\subsection{Femtoscopy}

Femtoscopic techniques, originally developed through the Hanbury Brown–Twiss (HBT)~\cite{PhysRevD.33.1314} method in astronomy, utilize two-particle momentum correlations to probe the space-time structure of the particle-emitting source at kinetic freeze-out. These measurements are primarily sensitive to final-state interactions, including Coulomb effects, strong final-state interactions (FSI), and quantum statistics. Nevertheless, they provide valuable constraints on the system size, lifetime, and collective expansion dynamics, offering complementary information to fluctuation observables in the exploration of the QCD phase structure.
\subsubsection{Identical pion femtoscopy}

\begin{figure}[ht!]
\centerline{\includegraphics[width=8cm]{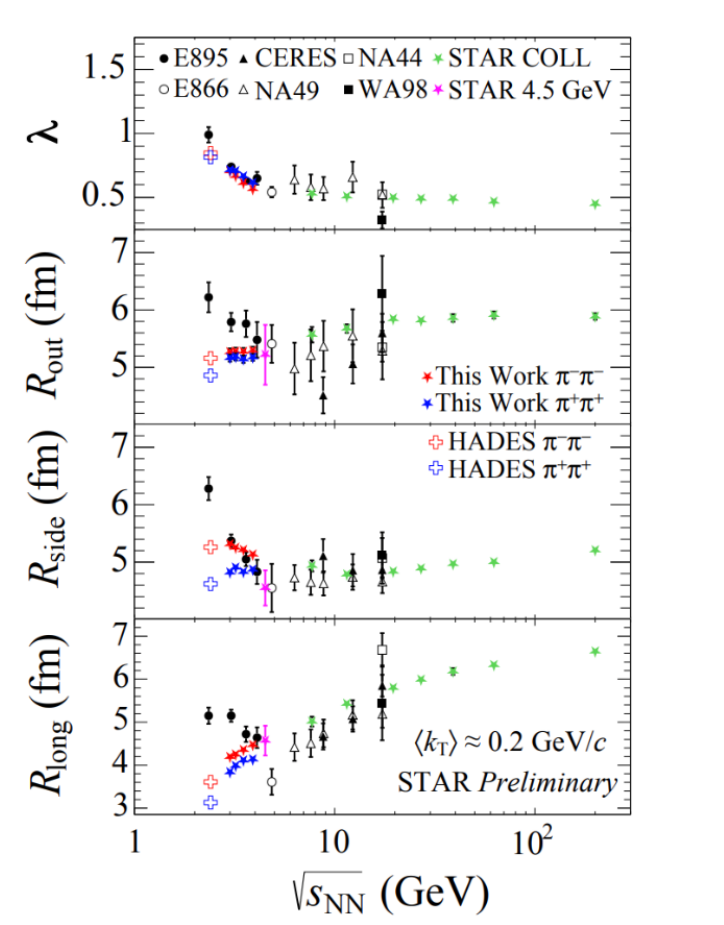}}
\caption{Collision-energy dependence of the femtoscopic radii extracted from identical-pion HBT correlation measurements in heavy-ion collisions. The radii characterize the spatial and temporal extent of the particle-emitting source at kinetic freeze-out. The measurements are compared with results from HADES and STAR over a broad range of $\sqrt{s_{NN}}$. Statistical uncertainties are shown as bars, while systematic uncertainties are represented by shaded boxes where applicable. Figure taken from Ref.~\cite{Luong:2024eaq}.\label{f4}}
\end{figure}

Figure~\ref{f4} presents the collision-energy dependence of the extracted femtoscopic radii obtained from identical charged pion  correlations in heavy-ion collisions. The femtoscopic parameters are measured over a broad range of collision energies, allowing a systematic study of the space--time evolution of the particle-emitting source at kinetic freeze-out.

The extracted radii exhibit trends that are consistent with previous measurements from HADES~\cite{HADES:2020ver} at lower collision energies and STAR at higher collision energies, providing a coherent picture of the evolution of the source geometry across the Beam Energy Scan program. In particular, a slight decrease of $R_{\mathrm{side}}$ and a gradual increase of $R_{\mathrm{long}}$ with increasing collision energy are observed. This behavior suggests a change in the shape and expansion dynamics of the emitting source as the system evolves from a more oblate geometry at low collision energies toward a more elongated, prolate configuration at higher energies.

The observed evolution of the femtoscopic radii reflects the interplay between transverse and longitudinal expansion dynamics and provides important constraints on the freeze-out conditions and the equation of state of the medium created in relativistic heavy-ion collisions.

\subsubsection{Hyperon-nucleon correlations}

\begin{figure}[ht!]
\centerline{\includegraphics[width=6.9cm]{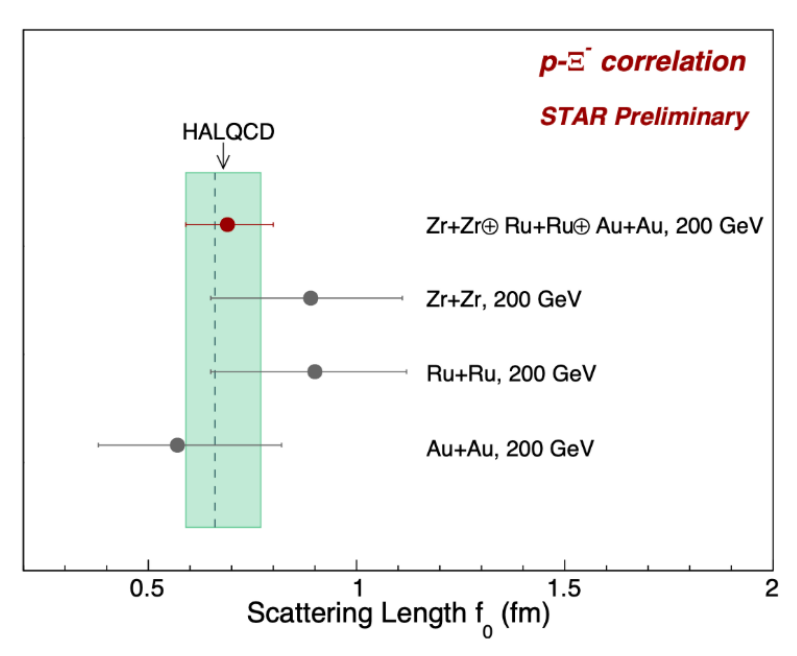}}
\caption{Extracted $p$--$\Xi^{-}$ scattering length parameter $f_{0}$ from femtoscopic correlation analyses in Au+Au, Ru+Ru, and Zr+Zr collisions at $\sqrt{s_{NN}} = 200$~GeV. The combined result from all collision systems is also shown and compared with theoretical predictions from HAL QCD calculations. Figure taken from Ref.~\cite{KZQM2025}\label{f5}}
\end{figure}

Figure~\ref{f5} shows the extracted $p$--$\Xi^{-}$ scattering length $f_{0}$ for different collision systems at $\sqrt{s_{NN}} = 200$~GeV. The measurements from Au+Au, Ru+Ru, and Zr+Zr collisions are found to be mutually consistent within uncertainties, indicating no strong system-size dependence of the extracted interaction parameters. The combined result from all collision systems provides an improved constraint on the $p$--$\Xi^{-}$ interaction.

The extracted scattering lengths are also compared with predictions from HAL QCD~\cite{Iritani_2019} calculations. Within uncertainties, the STAR measurements are consistent with the theoretical expectation, supporting the existence of an attractive $p$--$\Xi^{-}$ interaction. Such measurements are important for constraining hyperon--nucleon interactions, which play a crucial role in understanding the equation of state of dense baryonic matter and the composition of neutron stars.

The consistency of the extracted femtoscopic parameters across multiple collision systems further demonstrates the robustness of the femtoscopic analysis techniques and highlights the sensitivity of hyperon-nucleon correlations measurements to the underlying strong interaction potential.

\begin{figure}[ht!]
\centerline{\includegraphics[width=11.5cm]{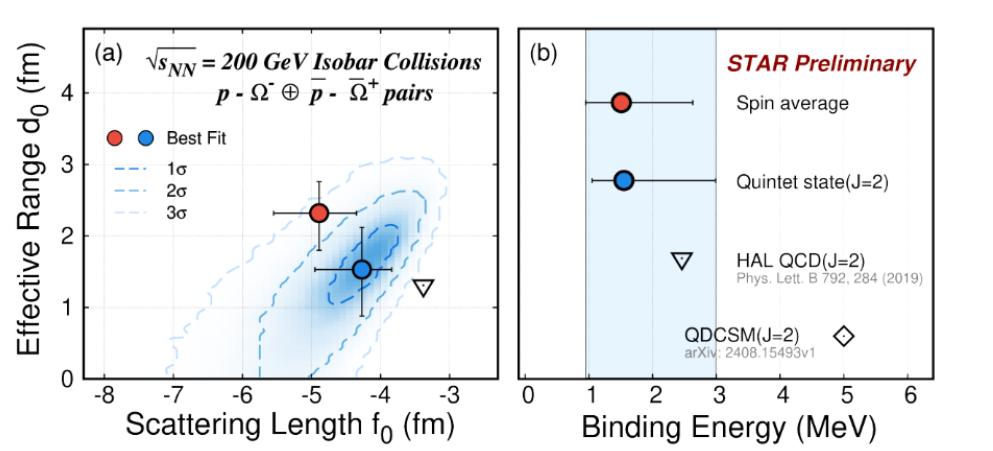}}
\caption{(a) Contour plot of the effective range $d_0$ versus scattering length $f_0$ for $p-\Omega^-$ and $\bar{p}-\Omega^+$ pairs measured in 200~GeV isobar collisions (Ru+Ru and Zr+Zr). The red and blue points indicate the best-fit values, with 1$\sigma$, 2$\sigma$, and 3$\sigma$ confidence contours shown. (b) Binding energy extracted for the spin-averaged and $J=2$ (quintet) states compared with theoretical predictions from HAL QCD and the Quark Delocalization Color Screening Model (QDCSM).  Figure taken from Ref.~\cite{KZQM2025}\label{f6}}
\end{figure}
The measurement of $p\text{--}\Omega^{-}$ and $\bar{p}\text{--}\bar{\Omega}^{+}$ correlation functions in isobar collisions at $\sqrt{s_{NN}} = 200$~GeV as shown in Figure~\ref{f6} provides new insights into the hyperon-nucleon interaction. By fitting the experimental correlation data, the scattering parameters—scattering length ($f_0$) and effective range ($d_0$)~\cite{Lednicky:1981su} were extracted to characterize the strong interaction potential.

The analysis yields an extracted negative scattering length ($f_{0}$), as illustrated in the $d_{0}$ vs. $f_{0}$ contour plot. The best-fit values  (indicated by the red and blue markers) are located in the negative $f_{0}$ region, specifically between $-4$ and $-5$ fm. The $1\sigma$, $2\sigma$, and $3\sigma$ confidence levels demonstrate a robust exclusion of positive $f_{0}$ values, confirming that the $p\text{--}\Omega$ interaction is predominantly attractive.

These results constitute the first experimental evidence of a strange dibaryon. The extracted scattering parameters are consistent with the existence of a bound state in the strangeness $S = -3$ sector. The data specifically probes the interaction in the spin-average and quintet ($J=2$) states, favoring a bound state scenario over a purely scattering one. 

Furthermore, the calculated binding energy is consistent with HAL QCD predictions~\cite{Iritani_2019} for the $J=2$ state. The experimental data points align closely with the HAL QCD prediction and the Quark Delocalization Color Screening Model (QDCSM)~\cite{yan2025investigatingpomegainteractioncorrelation}, with the binding energy concentrated in the $1$ to $3$~MeV range. These STAR Preliminary  results from isobar collisions at 200~GeV strongly support the hypothesis of a weakly bound $p\text{--}\Omega$ state. These femtoscopic measurements, spanning both the space-time evolution of the system and the detailed hadronic interaction potentials, provide complementary constraints on the QCD equation of state. When combined with fluctuation observables, they offer a more complete picture of the medium properties across the QCD phase diagram and help contextualize the ongoing search for the critical point.

\section{Summary}

The STAR experiment has presented a comprehensive set of new measurements from the Beam Energy Scan-II program, covering a wide range of collision energies from $\sqrt{s_{NN}} = 3$ to 200~GeV. These results include higher-order net-proton cumulants, two-particle transverse momentum correlations, baryon-strangeness correlations, and femtoscopic studies of both identical pions and baryon-baryon pairs.

The net-proton cumulant ratios show interesting energy dependence, with significant deviations from hadronic baselines observed at the lowest fixed-target energies. The two-particle transverse momentum correlator exhibits a non-monotonic behavior as a function of collision energy in central collisions, which is not fully reproduced by current transport models. Baryon-strangeness correlations display a clear transition from hadronic-like behavior at low energies toward partonic-like behavior at higher energies. Femtoscopic measurements provide detailed information on the space-time evolution of the system and yield new constraints on hyperon-nucleon interactions, including evidence supporting the existence of a weakly bound $p-\Omega$ state.

Collectively, these measurements provide complementary probes of the QCD medium across a broad range of the phase diagram. While no definitive signature of the critical point has been observed, the results place important new constraints on theoretical models and motivate continued high-precision measurements with the BES-II dataset and future fixed-target programs at even lower energies. Further theoretical developments and higher statistics data will be crucial in the ongoing search for the QCD critical point and the understanding of the QCD phase diagram.



\begingroup
\sloppy 
\bibliographystyle{ws-ijmpa}
\bibliography{sample}
\endgroup

\end{document}